\def\a{\alpha}\def\b{\beta}\def\c{\chi}\def\d{\delta}
\def\f{\phi}\def\g{\gamma}
\def\k{\kappa}\def\m{\mu}\def\q{\psi}\def\s{\sigma}
\def\y{\eta}\def\x{\xi}

\def\D{\Delta}

\def\de{\partial}
\def\inf{\infty}\def\id{\equiv}\def\mo{{-1}}

\def\tran{transformations }

\def\ct{conformal transformation }

\def\poi{Poincar\'e }

\def\SR{special relativity }

\def \schr{Schr\"odinger }

\def\section#1{\bigskip\noindent{\bf#1}\smallskip}

\def\PL#1{Phys.\ Lett.\ {\bf#1}}
\def\PRL#1{Phys.\ Rev.\ Lett.\ {\bf#1}}
\def\PR#1{Phys.\ Rev.\ {\bf#1}}\def\CQG#1{Class.\ Quantum Grav.\ {\bf#1}}

 \def\IJMP#1{Int.\ J. Mod.\ Phys.\ {\bf #1}}

\def\AoP#1{Ann.\ Phys.\ {\bf#1}}
\def\RMP#1{Rev.\ Mod.\ Phys.\ {\bf#1}}

\def\ref#1{\medskip\everypar={\hangindent 2\parindent}#1}
\def\beginref{\begingroup
\bigskip
\centerline{\bf References}
\nobreak\noindent}
\def\endref{\par\endgroup}

\baselineskip18pt\magnification=1200
\def\nrl {nonrelativistic limit }
\def\bp{{\bf p}}\def\bv{{\bf v}}
\def\yt{\tilde\y}\def\ct{\tilde\c}\def\ft{\tilde\f}\def\mt{\tilde m}
\def\umm{\left(1-{c^4m^2\over\k^2}\right)}
\def\ump{1+{c^2m\over\k}}\def\umn{1-{c^2m\over\k}}
\def\KG{Klein-Gordon }\def\ddr{deformed dispersion relation}
\def\dsr{deformed special relativity }
{\nopagenumbers
\line{\hfil May 2010}
\vskip40pt
\centerline{\bf The nonrelativistic limit of the Magueijo-Smolin model}
\centerline{\bf of deformed special relativity}
\vskip40pt
\centerline{
{\bf M. Coraddu}\footnote{$^\dagger$}{e-mail: massimo.coraddu@ca.infn.it}
and {\bf S. Mignemi}\footnote{$^\ddagger$}{e-mail: smignemi@unica.it}}
\vskip10pt
\centerline {Dipartimento di Matematica, Universit\`a di Cagliari}
\centerline{viale Merello 92, 09123 Cagliari, Italy}
\centerline{and INFN, Sezione di Cagliari, Cagliari, Italy}
\vskip60pt
\centerline{\bf Abstract}

\vskip10pt
{\noindent
We study the \nrl of the motion of a classical particle in a model of \dsr and of the
corresponding generalized \KG and Dirac equations, and show that they reproduce
nonrelativistic classical and quantum mechanics, respectively, although the rest mass
of a particle no longer coincides with its inertial mass.
This fact clarifies the meaning of the different definitions of velocity of a particle
available in DSR literature.
Moreover, the rest mass of particles and antiparticles differ, breaking the CPT
invariance. This effect is close to observational limits and future experiments may
give indications on its effective existence.}
\vfil\eject}

\section{1. Introduction.}
Classical mechanics can be recovered from relativistic mechanics through an expansion
in $1/c$, with the speed of light $c$ tending to infinite [1].
The same is true for quantum mechanics, where an analogous, but technically more
involved, expansion permits to obtain the \schr equation for a free particle from the
relativistic \KG and Dirac equations [2,3].

Although \SR describes very well the structure of spacetime at currently observable
energies, it has been argued that at energy scales close to the realm of quantum gravity,
i.e.\ near the Planck energy $\k=\sqrt{\hbar c^5\over G}=10^{19}$ GeV, \SR is deformed
in such a way that $\k$ becomes an observer-independent constant, like the speed of light
[4]. Models based on this assumption have been called deformed \SR (DSR),
and in their simplest form are introduced through a deformation of the relativistic
energy-momentum dispersion relation $E^2-c^2\bp^2=c^4m^2$, induced by the deformation of
the \poi invariance of the theory.
The deformation is required to be such that for $\k\to\inf$ one recovers special
relativity.

Given a DSR model, one may consider its \nrl for fixed $\k$, and
investigate if it correctly reproduces classical and quantum mechanics.
This limit corresponds to the speed of light
$c$ going to infinity, while the Planck constant $\hbar$ (that we shall set to 1 in
the following) and the Planck mass $\k/c^2$ are held fixed, and becomes relevant for
elementary particles of near-Planckian mass moving at low velocity, or for particles
at very low temperature. In the last case, in fact, the corrections due to deformations
of the dispersion relations are greater than those due to relativistic effects [5].
Although this limit is not experimentally observable at present,
it is interesting for checking the consistency
of the theory with the prediction of classical and quantum mechanics.

In this letter, we investigate the nonrelativistic limit of particle kinematics
in the case of the Magueijo-Smolin (MS) model [6], since this is algebraically the
simplest realization of DSR.
While the investigation of this limit is trivial in the classical case, it requires
the introduction of generalized \KG and Dirac equations in the quantum case.
We shall not study in detail the mathematical structure of the generalized
equations compatible with the \ddr, but will give an elementary
derivation of their nonrelativistic limit. At this level our derivation is
essentially independent of the position space realization of the MS model, which
we assume to be commutative, but depends only on the \ddr.

The result of our investigation is that the correct nonrelativistic limit is obtained.
However, the rest energy of the particle does not coincide with its inertial mass, but
contains corrections proportional to $c^2m/\kappa$.

Another important result is that the MS model induces a violation of the CPT symmetry,
resulting in a difference of the rest mass of particles and antiparticles, again of order
$c^2m/\kappa$. This feature is common to other realizations of DSR, as for example the
$\k$-\poi model [7], and originates from the breaking of the invariance of the dispersion
relations for $E\to-E$.
Although a complete discussion of this topic would require the study of the second
quantized theory, it seems that if a consistent second quantized theory exists, the
CPT violation is inescapable.

\section{2. The MS model}
The MS model [6] postulates a nonlinear action of the \poi group on momentum
space, such that, under a boost in the $x$ direction, the components of
the momentum transform as
$$E'={E\cosh\x+c\,p_1\sinh\x\over\D},\qquad
p'_1={p_1\cosh\x+E\sinh\x/c\over\D},$$
$$p'_2={p_2\over\D},\qquad\qquad p'_3={p_3\over\D},\eqno(1)$$
where
$$\D=1+{E(\cosh\x-1)+c\,p_1\sinh\x\over\k},\eqno(2)$$
with $\x$ the rapidity parameter. We have denoted with $E$ the energy, and with
$p_i$ the components of the 3-momentum $\bp$.
According to (1)-(2), the energy $E$ of a particle cannot exceed the Planck energy
$\k$ in any reference frame. For $\k\to\inf$, the \tran reduce to the usual Lorentz
transformations.

The transformations (1)-(2) leave invariant the quantity
$${E^2-c^2\bp^2\over(1-E/\k)^2}=c^4m^2,\eqno(3)$$
where $m$ is the so-called Casimir mass.

\section{3. Classical nonrelativistic limit.}
We recall that the \nrl of special relativity kinematics for a free particle
is obtained [1] starting from the dispersion relation
$$E^2-c^2\bp^2=c^4m^2,\eqno(4)$$
and expanding it for $\bp^2\ll c^2m^2$.
This yields
$$E=\sqrt{c^2\bp^2+c^4m^2}\sim c^2m+{\bp^2\over2m}+\dots\eqno(5)$$
The first term on the right hand side corresponds to the rest energy,
while the second reproduces the nonrelativistic kinetic energy.

In the MS model, one assumes instead the deformed dispersion relation (3).
Notice that, contrary to \SR, the dispersion relation is not invariant for
$E\to-E$.

Solving (3) for the energy, one obtains
$$E={-{c^4m^2\over\k}\pm\sqrt{\umm c^2\bp^2+c^4m^2}\over1-c^4m^2/\k^2}.
\eqno(6)$$
The correct sign for a positive-energy particle is the upper one.
Expanding as above, yields
$$E\sim{c^2m\over\ump}+{\bp^2\over2m}+\dots\eqno(7)$$
It results that the classical \nrl of the MS model coincides with that of
special relativity, except that the rest energy is given by
$$m^+={m\over\ump},\eqno(8)$$
and differs from the Casimir mass $m$. Moreover, the inertial mass can be
identified with the Casimir mass, whose physical significance was not
clear in the earlier literature.
The effect of the deformed dispersion relation (3) in the \nrl is
therefore simply a renormalization of the rest energy of the particle.

If one considers the lower sign in (6), one has at first order,
$$E\sim-\left({c^2m\over1-{c^2m\over\k}}+{\bp^2\over2m}+\dots\right).
\eqno(9)$$
The mass $m^-=-m/(1-c^2m/\k)$ will be identified with the rest mass of the
negative energy states in the Klein-Gordon and Dirac equations.

It may also be interesting to calculate the \nrl of the velocity of a
classical particle. In \SR the velocity can be defined in two equivalent ways
as the group velocity,
$$\bv_g={\de E\over\de\bp},\eqno(10)$$
or the phase velocity,
$$\bv_p={c^2\bp\over E}.\eqno(11)$$
In both cases its value is
$$\bv={\bp\over\sqrt{m^2+\bp^2/c^2}}\sim{\bp\over m}.\eqno(12)$$
It is well known that in DSR one obtains different results from the two
definitions, and this fact has raised a debate on the correct definition of the
velocity of a particle in these theories [8].

In particular, in the MS model, in view of (6),
$$\bv_g={\bp\over\sqrt{m^2+\umm\bp^2/c^2}}\sim{\bp\over m},\eqno(13)$$
while
$$\bv_p={\umm\bp\over-{c^2m^2\over\k}+\sqrt{m^2+\umm\bp^2/c^2}}
\sim\left(\ump\right){\bp\over m}={\bp\over m^+}.\eqno(14)$$
From the previous results, it appears that the identification of the the physical
velocity of the particle with the group velocity $\bv_g$ is more consistent than
the other definition with the nonrelativistic dynamics and with the identification
of $m$ with the inertial mass.

\bigbreak
\section{4. Nonrelativistic limit of the Klein-Gordon equation.}
We shall now consider the \nrl of the generalized Klein-Gordon equation
adapted to the MS model. We follow the exposition given for the standard
case in ref.\ [9].

In special relativity, the \nrl is obtained writing
the \KG equation in the form [2,9]
$$-\de_t^2\f=M^2\f,\eqno(15)$$
with $M=\sqrt{c^4m^2-c^2\de_k^2}$, and defining the fields
$$\f^\pm=\f\pm iM^\mo\de_t\f,\eqno(16)$$
that satisfy the equations
$$i\de_t\f^\pm=\pm M\f^\pm.\eqno(17)$$
Defining then the functions
$$\q^\pm={\rm exp}[\pm ic^2mt]\ \f^\pm,\eqno(18)$$
one obtains in the large $c$ limit,
$$i\de_t\q^\pm=\pm(M-mc^2)\q^\pm\sim\mp{1\over2m}\de_k^2\,\q^\pm,\eqno(19)$$
recovering the \schr equation for a particle of mass $m$.
\bigskip

In the MS realization of DSR, the generalized \KG equation can be obtained
performing the standard substitution
$E\to i\de_t$, $\bp\to-i\de_k$ in the dispersion relation (3).
In order to avoid problems with derivatives in the denominator, we
write it as
$$\left(-\de_t^2+c^2\de_k^2\right)\f=c^4m^2\left(1-{i\over\k}\,\de_t\right)^2\f
\eqno(20)$$
i.\ e.
$$\left[-\umm\de_t^2+2i\,{c^4m^2\over\k}\,\de_t+c^2\de_k^2\right]\f=c^4m^2\f.
\eqno(21)$$

Notice that in the massless case the \KG equation maintains the standard form.
Eq.\ (21) can be given the more compact form
$$L^2\f\id(i\de_t+a)^2\f=M^2\f,\eqno(22)$$
where we have defined
$$a={c^4m^2\over\k\umm},\qquad\qquad M={c^2\over\umm}\sqrt{m^2-\umm{\de_k^2
\over c^2}}.\eqno(23)$$

In order to obtain the \nrl one can proceed as in the standard case
discussed above. One defines the fields
$$\f^\pm=\f\pm M^\mo L\f,\eqno(24)$$
that obey the equations
$$L\f^\pm=\pm M\f^\pm.\eqno(25)$$
It is then easy to see that
$$L^2\f^\pm=M^2\f^\pm.\eqno(26)$$
One can now define the auxiliary fields
$$\ft^\pm={\rm exp}\,[ic^2m^\pm t]\,\f^\pm,\eqno(27)$$
where $m^\pm$ is the rest energy of the positive (negative) energy mode.
The auxiliary fields satisfy the equation
$$i\de_t\ft^\pm=[\pm M-c^2m^\pm-a]\,\ft^\pm.\eqno(28)$$
Expanding in $1/c$ one finally gets at first order in $1/c$,
$$i\de_t\ft^\pm\sim\mp{1\over2m}\,\de_k^2\,\ft^\pm.\eqno(29)$$
Hence, in the \nrl one obtains the usual \schr equation for a
particle of mass $m$, provided one subtracts the
rest energy $c^2m^\pm$.

Since the dispersion relation (3) is clearly not invariant under $E\to-E$,
in the Klein-Gordon equation the rest mass of the antiparticle does not coincide
with the rest mass of the particle, although their inertial masses are the same.
This may lead to instabilities in the case of second quantized
fields if no conservation laws prevent the decay.
Of course, our argument is only heuristic, and should be supported by a thorough
study of the second quantization of the theory.
Moreover, although this feature is common to several DSR models, it is
easy to construct DSR models that are invariant for $E\to-E$.

Experimental bounds on the mass difference between particles and antiparticles are
given in the case of the $K^0-\bar K^0$ system by $\D m_K/m_K\approx 10^{-18}$
[10]. This is consistent with our results, that predict a value of order
${c^2 m_K\over\k}\approx10^{-19}$. Moreover, they are sufficiently close to the
current experimental bounds to permit an experimental check in near future. Also,
observations may help to discriminate between different DSR models, which predict
different numerical values for the corrections.

\section{5. Nonrelativistic limit of the Dirac equation.}
Similar arguments can be used for the Dirac equation.
A generalization of the Dirac equation to the MS model can be obtained proceeding as
in the Lorentz case [3], seeking for a linear equation of the form
$$\left[E-c\,\a_kp_k-c^2m\left(\b-\d{E\over\k}\right)\right]\q=0,\eqno(30)$$
with $4\times4$ matrices $\a_k$, $\b$ and $\d$, which give rise to the dispersion
relation (3) when multiplied on the left by $E+c\,\a_kp_k+c^2m\left(\b-\d{E\over\k}
\right)$.

It is straightforward to check that this request implies the standard relations for the
anticommutators of the $\a_k$ and the $\b$, together with $\d^2=1$, $\{\a_k,\d\}=0$,
$\{\b,\d\}=1$, which permit to identify $\d$ with $\b$.

Putting as usual $E\to i\de_t$, $p_k\to-i\de_k$, one gets then
$$\left[i\de_t+c\,\a_k\de_k-c^2m\b\left(1-{i\over\k}\de_t\right)\right]\q=0.\eqno(31)$$
Multiplying on the left by $\b$, and setting $c=1$, the equation can also be written
in the "covariant" form
$$i\g^\m\de_\m\q-m\left(1-{i\over\k}\,\de_t\right)\q=0,\eqno(32)$$
where $\g^\m$ are the standard Dirac matrices. Also in this case, massless particles
obey the standard Dirac equation.

The nonrelativistic limit is obtained by writing equation (31) in a representation
in which the Dirac matrices assume the block form
$$\a^k=\pmatrix{0&\s_k\cr\s_k&0},\qquad\b=\pmatrix{1&0\cr0&-1},$$
and
$$\q=\pmatrix{\y\cr\c}.$$
with $\s_k$ the Pauli matrics and $\y$ and $\c$ column 2-vectors.

Defining time-dependent vectors
$$\q=\pmatrix{\yt\cr\ct}={\rm exp}\,[ic^2\mt t]\pmatrix{\y\cr\c},\eqno(33)$$
eq.\ (31) reduces to the system
$$i\left(1-{c^2m\over\k}\right)\yt-c\,\s_k\de_k\ct+c^2\left(\mt-m-{c^2m\mt\over\k}\right)\ft=0,
\eqno(34)$$
$$i\left(1+{c^2m\over\k}\right)\ct-c\,\s_k\de_k\ft+c^2\left(\mt+m+{c^2m\mt\over\k}\right)\ct=0.
\eqno(35)$$
Imposing that the last term of eq.\ (34) vanish, one obtains $\mt=m^+$, with $m^+$
given by (8).
Now, in the nonrelativistic limit, one can neglect the time derivative term in (35)
with respect to the other terms, obtaining
$$\ct\sim\left(\umn\right){\s_k\de_k\ft\over2m},\eqno(36)$$
and finally, substituting in (34) and using the properties of the Pauli matrices,
$$i\de_t\ft\sim{1\over2m}\,\de_k^2\,\ft,\eqno(37)$$
which is the \schr equation for a particle of mass $m$.

Imposing instead the vanishing of the last term of eq.\ (35), one obtains $\mt=m^-$,
and going through the same steps as before one obtains for the field $\ct$ describing
the antiparticle,
$$i\de_t\ct\sim-{1\over2m}\,\de_k^2\,\ct.\eqno(38)$$

As expected, the results of our analysis of the Dirac equation are analogous to those
obtained for the \KG equation, namely that particle and antiparticle have different
rest energies $m^\pm$, but identical inertial mass $m$.

\section{5. Final remarks}
Classically, DSR induces corrections of order ${c^2m/\k}$ on the rest mass of elementary
particles, while maintaining their inertial mass. Moreover, if the modified
dispersion relations are not invariant for $E\to-E$, particles and antiparticles
have different rest masses in the quantum theory.
In a second quantized theory, one may try to remedy this fact, and obtain a CPT
invariant model. However, it seems plausible that this cannot be done while maintaining
the classical limit (3) of the dispersion relation.
We expect that similar results hold also in the case of a noncommutative spacetime
realization of the theory (see f.e.\ [11] and references therein), although no detailed
study of this topic has been done.

The order of magnitude of the expected effects is close to the experimental limits.
Observations may therefore check the relevance of the theory and also discriminate
between different DSR models predicting different numerical values for the CPT
violations.

Finally, we remark that, even if the possible breakdown of CPT invariance is
model-dependent, the existence of a difference between rest and inertial masses
should hold for any DSR model.

\bigskip

\beginref
\ref [1] L.D. Landau, E.M. Lifshits, {\it The classical theory of fields},
Butterworth-Heinemann 1980.
\ref [2] H. Feschbach and F. Villars, \RMP{30}, 24 (1958).
\ref [3] J.D. Bjorken and S.D. Drell, {\it Relativistic quantum mechanics},
McGraw-Hill 1965.
\ref [4] G. Amelino-Camelia, \PL{B510}, 255 (2001), \IJMP{D11}, 35 (2002).
\ref [5] G. Amelino-Camelia, C. Laemmerzahl, F. Mercati and G.M. Tino, \PRL{103},
171302 (2009).
\ref [6] J. Magueijo and L. Smolin, \PRL{88}, 190403 (2002).
\ref [7] J. Lukierski, H. Ruegg and W.J. Zakrzewski, \AoP{243}, 90 (1995).
\ref [8] P. Kosi\'nski and P. Ma\'slanska, \PR{D68}, 067702 (2003);
S. Mignemi, \PL{A316}, 173 (2003);
M. Daszkiewicz, K. Imilkowska and J. Kowalski-Glikman, \PL{A323}, 345 (2004).
\ref [9] P. Bechouche, N. Mauser and S. Selberg, {\tt math}/0202201.
\ref [10] C. Amsler et al., \PL{B667}, 1 (2008).
\ref [11] A. Agostini, G. Amelino-Camelia and M. Arzano, \CQG{21}, 2179 (2004).

\endref
\end